\documentstyle[12pt,epsfig]{article}
\begin{document}
%

\def\etal{{\em et al.}}
\def\nue{\nu_{e}}
\def\nuebar{\overline{\nu}_{e}}
\def\numu{\nu_{\mu}}
\def\numubar{\overline{\nu}_{\mu}}
\def\nutau{\nu_{\tau}}
\def\nutaubar{\overline{\nu}_{\tau}}
\def\nux{\nu_x}
\def\nuxbar{\overline{\nu}_{x}}

\def\dmsq{\Delta m^{2}}
\def\sinsq{{\rm sin}^2 2\theta}
\def\Reines{$\overline{\nu}_e\,+\,p\,\rightarrow\,e^+\,+\,n$}

\def\SN#1E#2 {\mbox{$#1\times10^{#2}$}}      
\def\SNE#1+#2E#3 {\mbox{$(#1\pm#2)\times10^{#3}$}} 
\def\isotope#1{\mbox{${}^{#1}$}}                  
\def\ra{\rightarrow}
\def\lra{\leftrightarrow}
\def\units#1{\hbox{$\,{\rm #1}$}}                
%
\bigskip
\centerline{\Large\bf Initial Results from the CHOOZ Long Baseline}
\centerline{\Large\bf Reactor Neutrino Oscillation Experiment}
\begin{center}
M.~Apollonio$^c$,
A.~Baldini$^b$,
C.~Bemporad$^b$,
E.~Caffau$^c$,
F.~Cei$^b$,
Y.~D\'eclais$^e$,
H.~de~Kerret$^f$,
B.~Dieterle$^h$,
A.~Etenko$^d$,
J.~George$^h$,
G.~Giannini$^c$,
M.~Grassi$^b$,
Y.~Kozlov$^d$,
W.~Kropp$^g$,
D.~Kryn$^f$,
M.~Laiman$^e$,
C.E.~Lane$^a$,
B.~Lefi\`evre$^f$,
I.~Machulin$^d$,
A.~Martemyanov$^d$,
V.~Martemyanov$^d$,
L.~Mikaelyan$^d$,
D.~Nicol\`o$^b$,
M.~Obolensky$^f$,
R.~Pazzi$^b$,
G.~Pieri$^b$,
L.~Price$^g$,
S.~Riley$^g$,
R.~Reeder$^h$,
A.~Sabelnikov$^d$,
G.~Santin$^c$,
M.~Skorokhvatov$^d$,
H.~Sobel$^g$,
J.~Steele$^a$,
R.~Steinberg$^a$,
S.~Sukhotin$^d$,
S.~Tomshaw$^a$,
D.~Veron$^f$,
and V.~Vyrodov$^f$
\end{center}

\begin{center}
$^a${\em   Drexel University                    }  \\
$^b${\em   INFN and University of Pisa          }  \\
$^c${\em   INFN and University of Trieste       }  \\
$^d${\em   Kurchatov Institute                  }  \\
$^e${\em   LAPP-IN2P3-CNRS Annecy               }  \\
$^f${\em   PCC-IN2P3-CNRS Coll\`ege de France   }  \\
$^g${\em   University of California, Irvine     }  \\
$^h${\em   University of New Mexico, Albuquerque}  \\
\end{center}
\centerline{\large \sl PACS: \large 14.16.P, 28.50.Hw}

\begin{abstract}
\noindent
Initial results are presented from CHOOZ
\footnote{
The CHOOZ experiment is named after the new nuclear power station
operated by \'Electricit\'e de France (EdF) near the village of Chooz in
the Ardennes region of France.}, a long--baseline reactor--neutrino
vacuum--oscillation experiment. The data reported here were taken during
the period March to October 1997, when the two reactors ran at combined
power levels varying from zero to values approaching their full rated
power of 8.5\units{GW} (thermal). Electron antineutrinos from the
reactors were detected by a liquid scintillation calorimeter located at
a distance of about 1\units{km}. The detector was constructed in a
tunnel protected from cosmic rays by a 300\units{MWE} rock overburden.
This massive shielding strongly reduced potentially troublesome
backgrounds due to cosmic--ray muons, leading to a background rate of
about one event per day, more than an order of magnitude smaller than the
observed neutrino signal. From the statistical agreement between
detected and expected neutrino event rates, we find (at 90\% confidence
level) no evidence for neutrino oscillations in the $\nuebar$
disappearance mode for the parameter region given approximately by
$\dmsq > 0.9~10^{-3}\units{eV^2}$ for maximum mixing and $\sinsq > 0.18$ \ 
for large $\dmsq$.             

\end{abstract}



\section{Introduction}

In the widely accepted Standard Model of electroweak interactions,
neutrinos have zero rest mass. Since neutrino oscillation experiments
sensitively probe the existence of finite neutrino mass, such
experiments are important for testing and possibly improving the
Standard Model. These experiments are especially intriguing at present
because they yield a number of hints of significant neutrino oscillation
effects [1]. If validated, these hints imply that changes in the
Standard Model are required. Furthermore, a finite neutrino mass could
have significant consequences in many aspects of astrophysics and
cosmology.

The neutrino oscillation experiments measure neutrino fluxes from
greatly differing sources. Each of a number of experiments measuring
solar neutrino fluxes on earth always finds values smaller than the 
expected ones. Various interpretations of these results in terms of an
oscillation--induced disappearance of electron neutrinos have been
offered.

Several atmospheric (``cosmic--ray'') neutrino experiments find
$\numu/\nue$ ratios that are anom{\-}alously low by about a factor of
two, indicating a disappearance of $\numu$ or possibly an appearance of
$\nue$. 

These experimental results are often presented in the context of a model
with two neutrino eigenstates of mass $m_1$ and $m_2$ which mix to form
two flavour states. A pure beam of electron--flavoured neutrinos has a
survival probability which oscillates due to the $m_1-m_2$ mass
difference. For a single neutrino energy $E_\nu(\rm{MeV})$ and a
distance from the source $L$ (meters), the survival probability can be
written in terms of the mixing parameter $\sinsq$ and the difference of
the squared masses \linebreak[4] $\dmsq$ = $\left| {m_2^2-m_1^2}\right| $ as follows:

\begin{displaymath}
   P(\overline{\nu}_{e}\ra\overline{\nu}_{e}) =
   1 - {\rm  sin}^2 2\theta\ {\rm sin}^2
   \left(
   \frac{1.27\,\Delta m^{2}({\rm eV}^2)\,L({\rm m})} {E_\nu({\rm MeV})}
   \right).
\end{displaymath}

\noindent
When averaged over the source energy spectrum, this formula links the
disappearance of neutrinos to neutrino mass.

Due to their large values of $L/E$, the solar neutrino experiments 
favour $\dmsq$ values in the range $10^{-5}$ to $10^{-10}\units{eV^2}$,
while the cosmic--ray neutrino results give $10^{-1}$ to
$10^{-3}\units{eV^2}$. Ambiguities persist in the interpretation of
the cosmic--ray neutrino ratio; the $\numu$ flux may be low due to
either oscillation of $\numu \lra \nutau$ or to $\numu \lra \nue$.

The CHOOZ experiment [2] has an average value of $L/E\sim300$\ 
($L \sim1\units{km}$,\linebreak[4] $E\sim3 \units{MeV}$), 
an intense and nearly pure neutrino flavour
composition ($\sim100\%\ \nuebar$) and an intensity known to better than
$2\%$. It is thus ideally suited for a definitive test of $\nuebar \ra
\numubar$ neutrino oscillations (or, more generally, $\nuebar \ra \nuxbar$
oscillations) down to $10^{-3}\units{eV^2}$, an order of magnitude lower
than previous reactor experiments [3-7]. CHOOZ focuses directly on the
cause of the cosmic--ray anomaly and greatly increases the mass range
explored by neutrino experiments having well understood flux and
composition.

\section{Description of the CHOOZ Experiment}

\subsection{The Neutrino Source}

The Chooz power station has two pressurized water reactors with a total
thermal power of $8.5\units{GW_{th}}$. The first reactor reached full
power in May 1997, the second in August 1997. The $\nuebar$ flux and
energy spectrum of similar reactors have been extensively studied.
Analogous methods [3,8] have been used to calculate the $\nuebar$ flux
emitted by the Chooz reactors. The calculation includes

\begin{itemize}

\item a full description of the reactor core including the initial
\isotope{235}U enrichment and the daily evolution [9] of the isotopic
composition of each fuel element as a function of the power
produced (burn--up),

\item the instantaneous fission rate derived from the thermal power of the
reactors (recorded every minute), 

\item the $\nuebar$ yield (as determined in [10--12]) from the four main
isotopes -- \isotope{235}U, \isotope{238}U, \isotope{239}Pu, and
\isotope{241}Pu.

\end{itemize}

It has been shown [3] that the value of the $\nuebar$ flux
emitted by a reactor is understood to 1.4\%. For this reason there is no
need for a normalization experiment to measure the $\nuebar$ flux with a
detector close to the reactors.

\subsection{The Site}

The detector is located in an underground laboratory at a distance of
about 1\units{km} from the neutrino source. The 300\units{MWE} rock
overburden reduces the external cosmic ray muon flux by a factor of
$\sim300$ to a value of $0.4\units{m^{-2}s^{-1}}$, significantly
decreasing the most dangerous background, which is caused by fast
neutrons produced by muon--induced nuclear spallation in the materials
surrounding the detector. This cosmic ray shielding is an important
feature of the CHOOZ site.

The detector (Fig.~1) is installed in a welded cylindrical steel vessel
5.5\units{m} in diameter and 5.5\units{m} deep. 
\begin{figure}[htb]
\begin{center}
\mbox{\epsfig{file=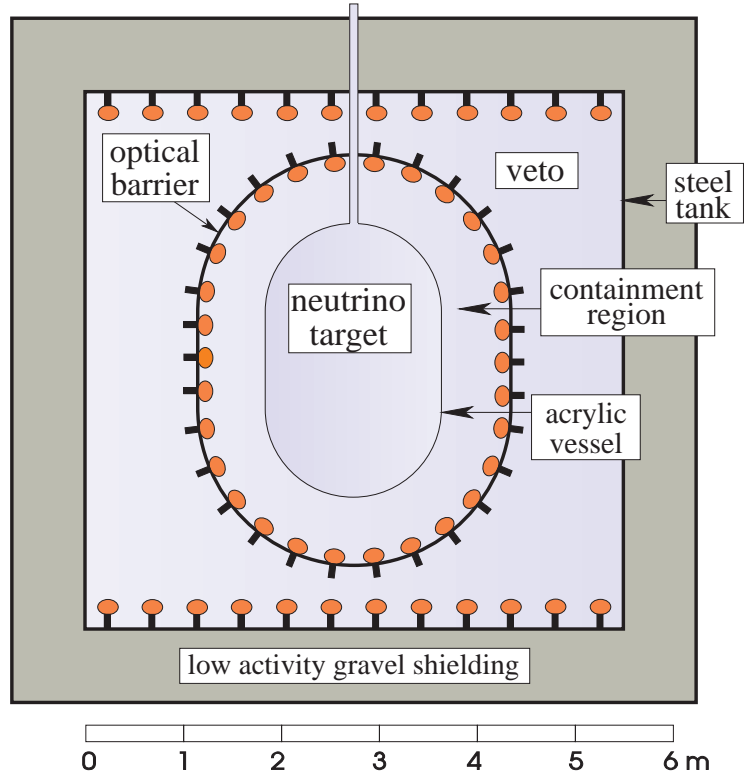,width=0.70\textwidth,%
bbllx=65bp,bblly=130bp,bburx=530bp,bbury=650bp}}
\caption{The CHOOZ detector.}
\end{center}
\end{figure}
The inside walls of the
vessel are painted with high--reflectivity white paint. The vessel is
placed in a pit 7\units{m} in diameter and 7\units{m} deep. To protect
the detector from the natural radioactivity of the rock, the steel
vessel is surrounded by 75\units{cm} of low radioactivity sand
(Comblanchien from Burgundy in France) and covered by 14\units{cm} of
cast iron.

\subsection{The Detector}

The $\nuebar$
are detected via the inverse beta decay reaction
$$\displaylines{
 \overline{\nu}_e\,+\,p\,\rightarrow\,e^+\,+\,n\ {\quad\rm with\quad}
 E_{e^+}=E_{\overline{\nu}_e}-1.804\,{\rm MeV} \ .              \cr
}$$

\noindent
The $\nuebar$ signature is a delayed coincidence between the prompt
$e^+$ signal (boosted by the two 511--\units{keV} annihilation gamma rays)
and the signal from the neutron capture. The target material is a
hydrogen--rich (free protons) paraffinic liquid scintillator loaded with 0.09\%
gadolinium. The target is contained in an acrylic vessel of precisely
known volume immersed in a low energy calorimeter made of unloaded
liquid scintillator. Gd has been chosen due to its large neutron capture
cross section and to the high $\gamma$-ray energy released after n-capture 
($\sim8\units{MeV}$, well above the natural radioactivity).

The detector is made of three concentric regions:

\begin{itemize}

\item a central 5--ton target in a transparent plexiglass container
filled with a 0.09\% Gd--loaded scintillator (``region 1'');

\item an intermediate 17--ton region (70\units{cm} thick) equipped with 192
eight--inch PMT's (15\% surface coverage, $\sim 130$ photoelectrons/MeV) 
\ [13], used to protect the target from PMT radioactivity and 
to contain the gamma rays from neutron capture (``region 2'');

\item an outer 90--ton optically separated active cosmic--ray muon veto
shield (80\units{cm} thick) equipped with two rings of 24 eight--inch PMT's
 (``region 3'').

\end{itemize}

The detector is simple and easily calibrated, while its behaviour can be
well checked. Six laser flashers are installed in the three regions
together with calibration pipes to allow the introduction of radioactive
sources. The detector can be reliably simulated by the Montecarlo
method [14].

The first target filling needed to be replaced after about four months
of operation due to a problem with the Gd--loaded scintillator ---
reduced transparency of the scintillator caused by a chemical
instability. After protecting with teflon all metallic parts in contact
with the liquid and after slightly modifying the scintillator formula
for improved stability, we now operate with a new filling introduced in
March 1997. The new scintillator has a measured useful lifetime in the
detector (due to reduced light at the PMT's) of $\tau\sim750\units{d}$.

The period with the first scintillator fill was very useful in tuning
the detector and in establishing the appropriate calibration methods and
analysis procedures. All data used in the present paper were gathered
using the second batch of scintillator.

Fig.~2 shows calibration data using a \isotope{252}Cf source at the
detector centre. 
\begin{figure}[t]
\begin{center}
\mbox{\epsfig{file=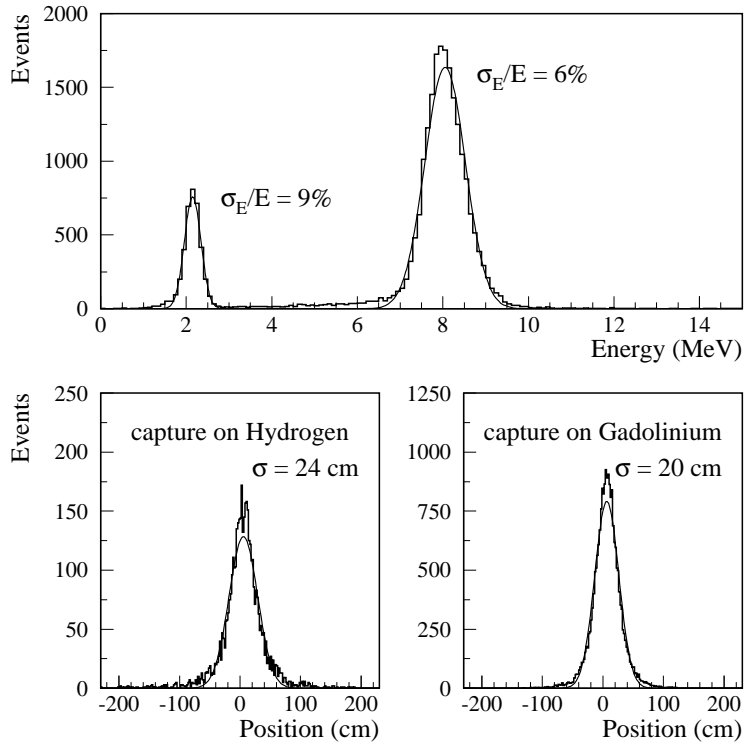,width=0.70\textwidth,%
bbllx=20bp,bblly=130bp,bburx=540bp,bbury=665bp}}
\caption{Energy and position resolution for \  $^{252}$Cf $n$-capture 
$\gamma$-lines.}
\end{center}
\end{figure}
The neutron capture lines (2.2\units{MeV} on hydrogen
and 8\units{MeV} on gadolinium) are clean and the energy resolutions are
$\sigma_E/E=9\%$ and $\sigma_E/E=6\%$, respectively.

\subsection{Trigger and Acquisition Electronics}

The experiment is triggered using a two--level, two--threshold scheme.
Level--one triggers, formed by discriminating the total PMT charge and
hit multiplicity, correspond to deposited energies of 1.3\units{MeV}
(L1$_{low}$, rate $\sim130\units{s^{-1}})$ and 3.3\units{MeV}
(L1$_{high}$, rate $\sim30\units{s^{-1}})$. The level--two trigger
(L2) requires a delayed coincidence of L1$_{low}$ and L1$_{high}$
within $100\units{\mu s}$. Both pulse sequences are allowed in order
to permit study of the accidental background. Finally, the readout of
neutrino candidate events (rate $\sim0.15\units{s^{-1}}$) is triggered
by an L2 preceded by at least a 1\units{ms} time interval with no
activity in the veto shield.

The data acquisition system consists of several redundant digitization
systems controlled by a VME--based processor. Each PMT signal is
digitized by a multi--hit TDC and by one of two alternating ADC banks,
gated by L1$_{low}$. Signals from groups of 8 PMT's, referred to as {\em
patches}, are fanned--in and recorded by fast (150 MHz, $8\times206$ ns
deep) and slow (20~MHz, 200\units{\mu s} deep) waveform digitizers and
by two banks of multi--hit VME ADC's gated by L1$_{low}$. The latter
provide charge information for a total of nine L1$_{low}$ events. The
VME ADC's are also connected with a neural--network--based event
reconstruction fast processor (NNP) [15]. The NNP reconstructs the
position and the energy of events in $\sim150\units{\mu s}$, allowing
study of events corresponding to much lower energy thresholds than those
permitted by the main CHOOZ trigger.

\section{Data Analysis}

\subsection{Event Selection and Reconstruction}

The redundancy in the digitization electronics allows a good check of
data quality. It has been found that using the 8:1 fanned--in patches
results in good energy and position resolutions. The results presented
in this paper are based on the 8:1 fanned--in VME ADC data, fitted by a
MINUIT--based procedure to obtain energy and position. Procedures using
other charge and time information are also being studied.

The selection of events is based on the following requirements:

\begin{itemize}
\item energy cuts on the neutron candidate ($6-12$ MeV) and on the
$e^+$ (from the L1$_{low}$ threshold energy to 8 MeV),

\item a time cut on the delay between the $e^+$ and the
neutron ($2-100\units{\mu s}$), 

\item spatial cuts on the $e^+$ and the neutron (distance from the PMT
wall $> 30\units{cm}$ and distance $n-e^+ < 100\units{cm}$).
\end{itemize}

The events are then divided in two classes. The first class contains the
neutrino candidates, in which the $e^+$ precedes the $n$, while the
second consists of events in which the $n$ precedes the $e^+$. The
second class of events is used to establish the rate of accidentals and
of multiple neutron events. These events provide a good test of the
stability of the background throughout the data taking periods.
\begin{figure}[t]
\begin{center}
\mbox{\epsfig{file=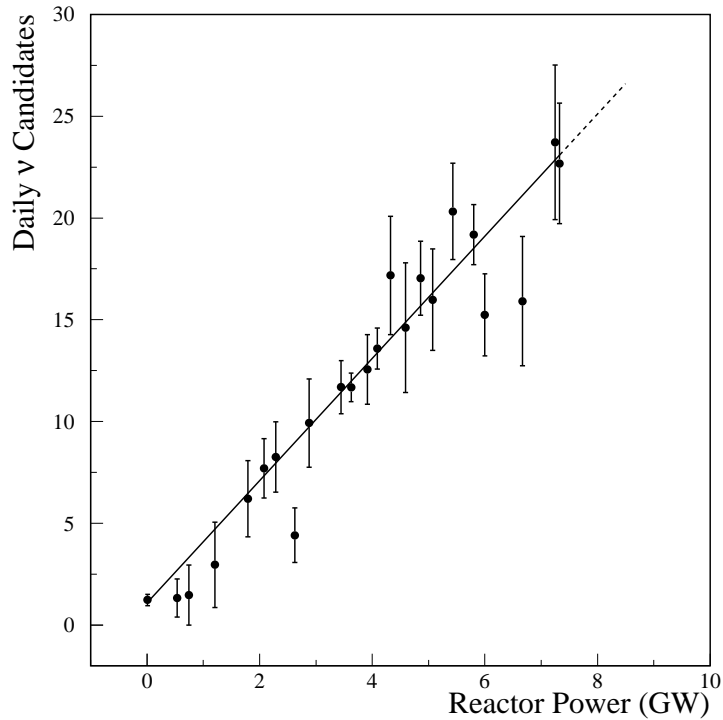,width=0.65\textwidth,%
bbllx=40bp,bblly=140bp,bburx=540bp,bbury=665bp}}
\caption{Number of \ {$\nuebar$}-candidates d$^{-1}$, as a function 
of the reactor power.} 
\end{center}
\end{figure}

\subsection{Background and Neutrino Signal}

The neutrino signal was obtained by subtracting the small number of
accidental and correlated background events which passed the neutrino
selection criteria. These two components were determined separately.
The accidental component was estimated by using the $e^+$ and $n$
uncorrelated rates. The correlated component was estimated from
reactor--off data by extrapolating the rate of high--energy neutrons
followed by $n$-capture into the region defined by the $\nu$ event
selection criteria. The background rates were found to be

\begin{displaymath}
    R_{accidental}        = 0.23\pm0.05\units{d^{-1}}\ {\rm and} \
    R_{correlated}        = 0.8 \pm0.2 \units{d^{-1}}.
\end{displaymath}

The total background rate was measured with reactor-off 
and also by extrapolating the
neutrino candidate rate to zero reactor power. For this purpose,
neutrino candidates were collected during the period of reactor power
rise. The number of events corresponding to each data run depended on
the run duration $\Delta T$ and on the average reactor power $W_{av}$.
All data were therefore fitted as a function of these two variables,
thus separately determining the neutrino signal and the reactor--off
background. The (grouped) data are shown in Fig.~3 as a function of
reactor power.
The superimposed line corresponds to the fitted signal
and background values. The results of the various background
determinations are summarized in Table 1.

\bigskip
\centerline{\bf \qquad Table 1.\quad Background Rates}
\bigskip

\centerline{
\begin{tabular}{|l|c|} \hline
estimated rate                      & $1.03\pm0.21\units{d^{-1}} $ \\[2mm] \hline
reactor--off rate                   & $1.2 \pm0.3 \units{d^{-1}} $ \\[2mm] \hline
rate by extrapolation to zero power & $1.1\pm0.25 \units{d^{-1}} $ \\[2mm] \hline
\end{tabular}
            }
\bigskip

It should be noted that the measurements of the background are in good
agreement. From the fit we find the neutrino rate $S_{fit}$ normalized
to the full power of the two reactors ($2\times 4.25\units{GW_{th}}$,
$2\times \SN1.3E20 \units{fissions\ s^{-1}}$) and a burn--up of
$1300\units{MW\, d \ ton^{-1}}$ to be

\begin{displaymath} S_{fit}=25.5\pm1.0\units{d^{-1}}. \end{displaymath}

A total of 1320 neutrino events were accumulated during 2718 hours (live
time). The distribution of their physical parameters (neutron capture
energy and time, $n-e^+$ distance) is presented in Fig.~4; a neutron 
delay distribution 
obtained by a \isotope{252}Cf source is also shown.
\begin{figure}[bp]
\begin{center}
\mbox{\epsfig{file=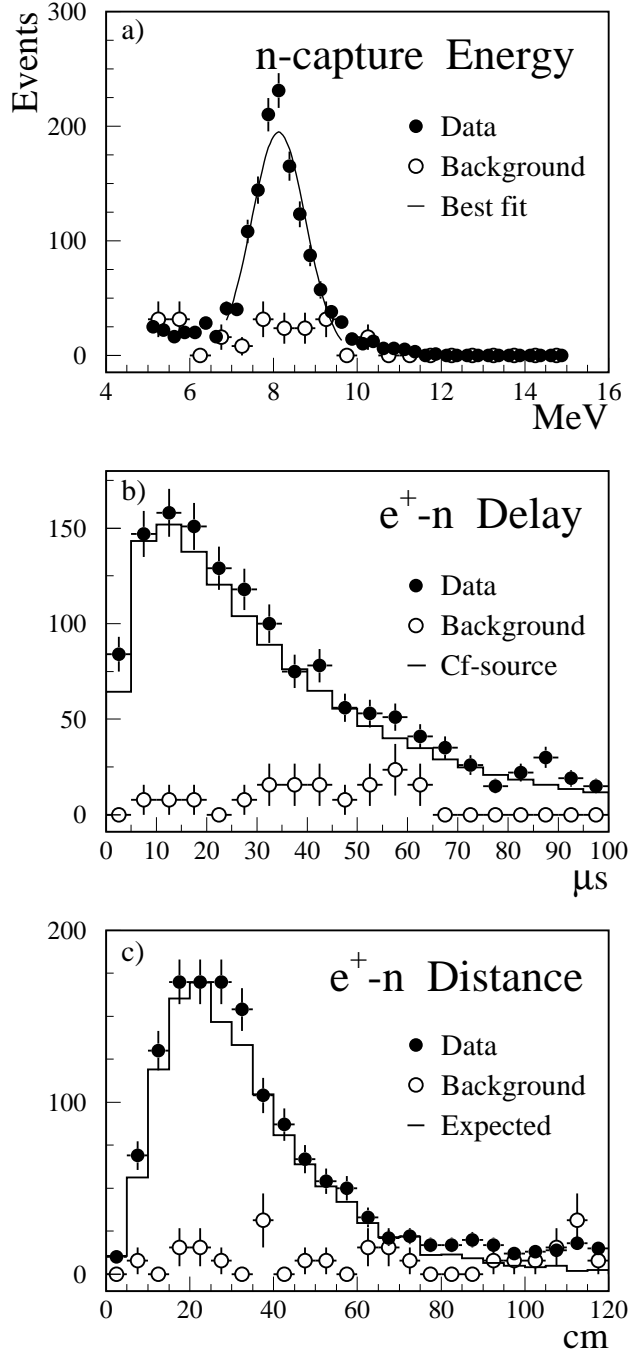,width=0.60\textwidth,%
bbllx=105bp,bblly=35bp,bburx=465bp,bbury=775bp}}
\caption{Distribution of: \ a)~energy released by $n$-capture on Gd,
\ b)~$n$-capture delay,\ c)~positron-neutron distance, measured and MC 
expected; the reactor-off background distribution is also shown.
The histograms in \ b) and \ c) \ are normalized to the background 
subtracted experimental data.} 
\end{center}
\end{figure}
Using the total number of events and subtracting the reactor--off
background, we obtain the similar, but slightly less accurate, result

\begin{displaymath}S_{integral}=25.0\pm1.1\units{d^{-1}} . \end{displaymath}

\section{Expected Number of Events and Systematic Errors}

The expected number of detected neutrino events can be written

\begin{displaymath}
N_{ev}=N_{fiss}
\times \sigma_{fiss}
\times \frac{1}{4\pi D^2}
\times n_p
\times \epsilon_{e^+}
\times \epsilon_n
\times \epsilon_{\Delta r}
\times T_{live}\ ,
\end{displaymath}

\noindent where

\begin{itemize}

\vspace{-0.2cm}
\item $N_{fiss}$ is the number of fissions, per unit time, 
in the reactor cores,

\vspace{-0.2cm}
\item $\sigma_{fiss}$ is the cross section [4,16,17]
for the reaction~~\Reines
~~cal\-culated for a neutron lifetime of $887.0\pm2.0\units{s}$ [18],
and integrated over the $\nuebar$ energy spectrum and the fuel composition
of the core,

\vspace{-0.2cm}
\item $D$ is the distance of the detector from the reactors,

\vspace{-0.2cm}
\item $n_p$ is the effective number of free protons in the target,
including small corrections for edge effects involving the acrylic
vessel and the region 2 scintillator buffer,

\vspace{-0.2cm}
\item $\epsilon_{e^+}$ and $\epsilon_n$ are the positron and neutron detection
efficiencies averaged over the entire sensitive target,

\vspace{-0.2cm}
\item $\epsilon_{\Delta r}$ is the efficiency of the $e^+-n$ distance cut, and

\vspace{-0.2cm}
\item $T_{live}$ is the live time.

\end{itemize}

\noindent
Table 2 shows the various parameters together with their errors. The
combined systematic error is estimated at $\sim4\%$.

\bigskip
\centerline{\bf Table 2.\quad Normalization Parameters and Errors}
\bigskip

\centerline{
\begin{tabular}{|l|c|r|} \hline
{\bf \qquad Parameter}                                   &  {\bf Value}              & {\bf Error}  \\[2mm] \hline
cross section per fission                                & \SN6.327E-43 \units{cm^2} & 2.7\%        \\[2mm] \hline
reactor 1 distance                                       & \SN1.1146E5 \units{cm}    & 10\units{cm} \\
reactor 2 distance                                       & \SN0.9979E5 \units{cm}    & 10\units{cm} \\[2mm] \hline
number of free protons in the Gd--loaded scintillator    & \SN3.601E29               & 1.5\%        \\
effective number of free protons in the target           & \SN3.637E29               & 1.5\%        \\[2mm] \hline
$e^+$ efficiency                                         & 0.968                     & 0.7\%        \\[2mm] \hline
measured $n$ efficiency at centre (\isotope{252}Cf data) & 0.757                     & 1.3\%        \\
calculated $n$ efficiency at centre (Monte Carlo)        & 0.759                     & 1.3\%        \\
$n$ efficiency averaged over the effective volume        & 0.739                     & 2.0\%        \\[2mm] \hline
efficiency of the $e^+-n$ distance cut                   & 0.965                     & 1.4\%        \\[2mm] \hline
\end{tabular}
            }

\section{Results}

The ratio of the measured to expected neutrino signal is

\begin{displaymath}
R_{measured/expected}=0.98\pm0.04({\rm stat})\pm0.04({\rm syst}).
\end{displaymath}

\noindent
One can also compare the measured positron spectrum with what is
expected in the case of no oscillations (Fig.~5).
\begin{figure}[p]
\begin{center}
\mbox{\epsfig{file=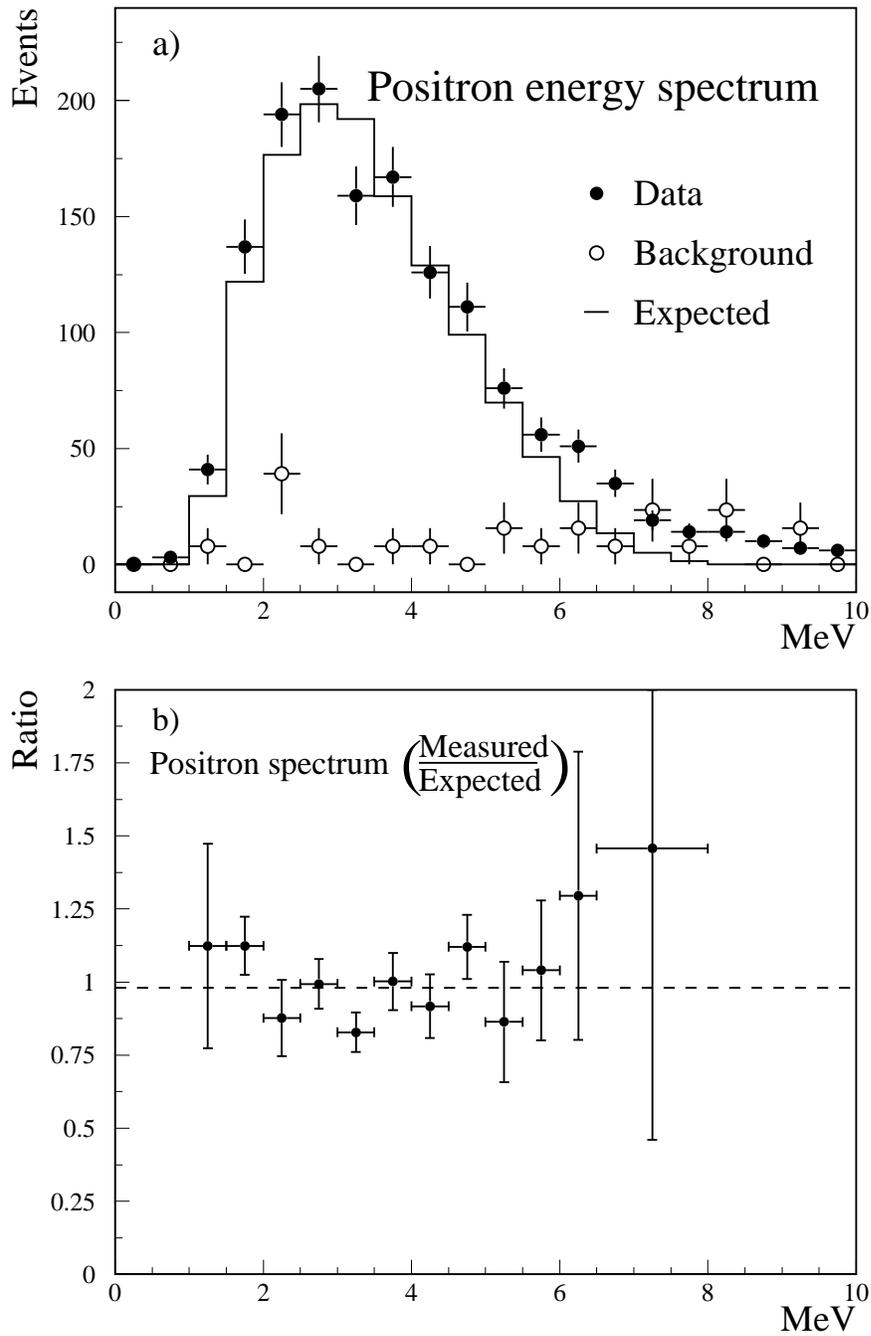,width=0.80\textwidth,%
bbllx=40bp,bblly=35bp,bburx=525bp,bbury=775bp}}
\caption{a)~Positron energy spectrum and corresponding reactor-off 
background for the same live-time;\ the neutrino-signal expected positron 
spectrum is also shown. b)\ Ratio of the measured (background subtracted) 
to the expected positron spectrum.} 
\end{center}
\end{figure}
The 90\% C.L.exclusion plot is
presented in Fig.~6, together with the exclusion plots of previous
experiments and the atmospheric neutrino signal reported by Kamiokande [19].
\begin{figure}[p]
\begin{center}
\mbox{\epsfig{file=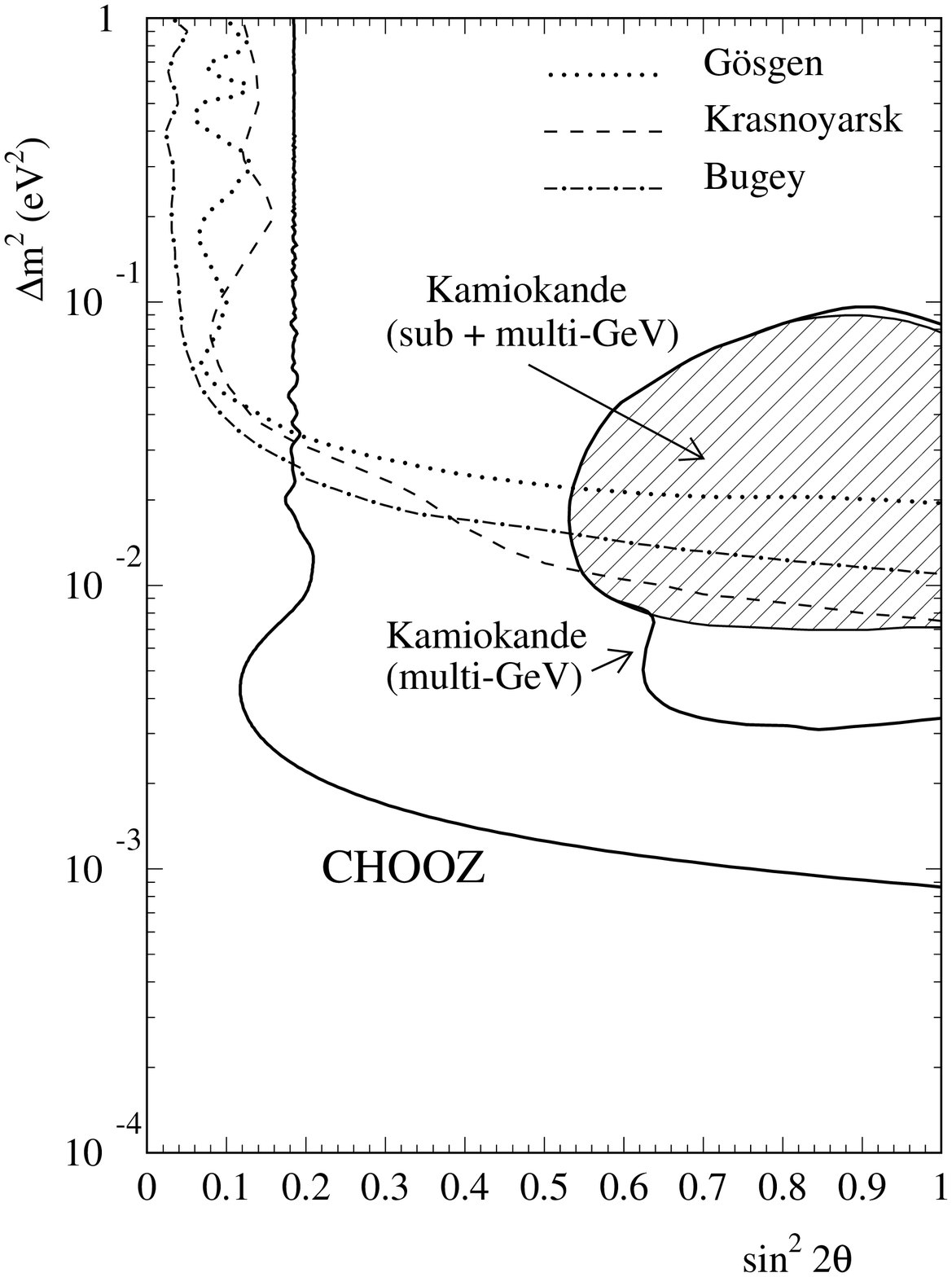,width=0.80\textwidth,%
bbllx=10bp,bblly=60bp,bburx=510bp,bbury=730bp}}
\caption{The 90\% C.L. exclusion plot for CHOOZ, 
compared with previous experimental 
limits and with the KAMIOKANDE allowed region.} 
\end{center}
\end{figure}

\section{Conclusions}

The CHOOZ experiment finds, at 90\% C.L., no evidence for neutrino
oscillations in the disappearance mode
$\overline{\nu}_{e}\ra\overline{\nu}_{x}$ for the parameter region given
approximately by $\dmsq > 0.9~10^{-3}\units{eV^2}$ for maximum mixing and 
$\sinsq > 0.18$ for large $\dmsq$, as shown in Fig. 6.

The experiment is continuing to take data in order to achieve better
statistics and to improve understanding of systematic effects.

\section{Acknowledgements}

Construction of the laboratory was funded by \'Electricit\'e de France
(EdF). Other work was supported in part by IN2P3--CNRS (France), INFN
(Italy), the United States Department of Energy, and by RFBR (Russia).
We are very grateful to the Conseil G\'en\'eral des Ardennes for having
provided us with the headquarters building for the experiment. At
various stages during construction and running of the experiment, we
benefited from the efficient work of personnel from SENA (Soci\'et\'e
Electronucl\'eaire des Ardennes) and from the EdF Chooz B nuclear plant.
Special thanks to the technical staff of our laboratories for their
excellent work in designing and building the detector.

\end{document}